\documentstyle[11pt,aaspp4]{article}

\slugcomment{Submitted Ap. J. Lett.}

\begin{document}

\title{Pulse Arrival-Times from Binary Pulsars with\\
       Rotating Black Hole Companions}

\author{Pablo Laguna\altaffilmark{1} and Alex Wolszczan\altaffilmark{2}}
\affil{Department of Astronomy \& Astrophysics and\\
       Center for Gravitational Physics \& Geometry\\
       Penn State University, University Park, PA 16802}

\altaffiltext{1}{pablo@astro.psu.edu}
\altaffiltext{2}{alex@astro.psu.edu}

\begin{abstract}
We present a study of the gravitational time delay of arrival of
signals from binary pulsar systems with rotating black hole companions.
In particular, we investigate the strength of this effect (Shapiro delay)
as a function of the inclination, eccentricity and period of the orbit,
as well as the mass and angular momentum of the black hole.
This study is based on direct numerical integration of null geodesics
in a Kerr background geometry.
We find that, for binaries with sufficiently high orbital inclinations
($> 89^o$) and compact companion masses $> 10 M_\odot$,
the effect arising from the rotation of the black hole in the system
amounts to a microsecond--level
variation of the arrival times of the pulsar pulses.
If measurable, this variation could provide a unique signature for the
presence of a rotating black hole in a binary pulsar system.
\end{abstract}


Binary pulsars are excellent laboratories for testing General Relativity (GR)
and other theories of gravity. Long-term timing observations of relativistic
binaries, such as PSR B1913+16 (\cite{hulse75}), yield highly
precise measurements of post-Keplerian orbital parameters of the pulsar.
In particular, observations of the the orbital decay of PSR B1913+16
induced by gravitational radiative effects have verified
the validity of GR to within $0.5\%$ (\cite{taylor89}).
Recently, pulsar surveys have unveiled additional relativistic binary pulsars
(for a recent review see Wolszczan (1997)).
Of particular interest is PSR B1534+12 (\cite{alex91,taylor92}),
in which case the presence of post--Keplerian orbital effects and
proximity of the pulsar to Earth give this object a potential to become
an exceptionally precise probe of relativistic gravity.

One measurable relativistic effect in the timing of binary pulsars is
the time delay that pulses experience as a result of the space-time curvature
induced by the pulsar's companion.
This is precisely the effect proposed by Shapiro (1964)
as a fourth test of GR, in addition to the tests
to measure perihelion shift, deflection of light and gravitational redshift.
The Shapiro delay is measurable in binary pulsars with
high orbital inclinations. So far, it has been detected in two low--mass
binaries, PSR B1855+09 (\cite{kaspi94}; $i \approx 87.7^o$) and
PSR J1713+07 (\cite{camilo94}; $i \approx 78.5^o$), and
in the high--mass PSR B1534+12 system mentioned above ($i \approx 78.4^o$).
The maximum amplitude of the pulse-timing residuals from the Shapiro delay
for these binary pulsars is approximately 15.0, 7.5, and 55.0 $\mu$s,
respectively.

The relativistic orbital effects in compact
neutron star--neutron star or neutron star--white dwarf binaries
are accounted for in a theory-independent timing formula
developed by Damour and Deruelle (1985, 1986). Since any gravitational effects
caused by the rotation of companions to pulsars in such binaries are
practically negligible,
this timing model does not take them into account.
Nonetheless, future pulsar detections could yield binary pulsars with
black hole (BH) companions, thereby creating
favorable conditions to measure the impact of the companion's rotation
on the timing of pulses.  In particular, Monte Carlo
simulations by Lipunov et al. (1994) indicate that, in the best case
scenario, the number of PSR--BH binaries is expected to be $\sim 1$ per 700 pulsars,
with BH masses of $\sim 20 M_\odot$.
If rotational effects are included, then, as shown below,
the inertial frame drag by the gravitational field
produces both an extra time delay and an advance of signals,
depending on the relative orientation of the orbital
angular momentum of the pulsar and the rotational angular momentum of the
companion.

In this Letter, we investigate the conditions necessary to
detect the effects of companion rotation
on the pulse arrival-times from a binary pulsar.
In particular, we are concerned with a possible application of such
measurements to gather additional information regarding the nature of the
pulsar's companion.
Although it is possible to derive a timing formula for the Shapiro delay
without a specific theory of gravity in mind, our results are obtained
using the theory of GR.  We start by considering the
space-time curvature in the vicinity of the companion.
If rotation is ignored, Birkhoff's theorem (\cite{hawking73})
guarantees that the only solution describing the exterior
gravitational field of an object is the Schwarzschild metric.
There is, however, no such theorem for the space-time geometry
exterior to a rotating object. Nonetheless, it is expected
that for a ``clean" binary pulsar, given enough time,
the gravitational field of a rotating companion will
settle down to the Kerr family of stationary and axisymmetric
space-times (\cite{wald84}). Our study then assumes that
the propagation of the pulses takes place in a Kerr space-time background;
that is, we consider PSR--BH binary systems.

Before addressing the problem in its full generality,
we consider, as an illustration,
the case of a gravitational time delay of signals
propagating in the equatorial plane of a rotating black hole.
From integrating the null geodesic equation for the Kerr
metric, one obtains that the time required for light to travel
in the equatorial plane from an arbitrary point $r$ to the distance
$d$ of closest approach ($d \ll r$) is given by
(\cite{dymnikova84,dymnikova86})
\begin{equation}
t^\pm(r,d) = \sqrt{r^2-d^2}
+ 2M_h\ln{\left(\frac{r+\sqrt{r^2-d^2}}{d}\right)}
+ M_h\left(\frac{r-d}{r+d}\right)^{1/2}
+\frac{(15\pi-8)M_h^2}{4d} \mp \frac{4aM_h}{d}\, ,
\label{eq:1}
\end{equation}
with $M_h$ and $a$ the mass and angular momentum per unit mass of
the black hole, respectively.
In the last term, the upper (lower) sign is chosen if the
orbital angular momentum of the photon is (anti-)parallel
to the angular momentum of the black hole.
Because of this relative sign,
the photon is accelerated, if its motion
is along the direction of the black hole rotation and
it is delayed otherwise.
The first term in equation [\ref{eq:1}] corresponds to
signal propagation in a flat space-time.
The second and third terms encapsulate the Shapiro
delay as originally proposed (\cite{shapiro64}).
The fourth term in equation [\ref{eq:1}]
is the second order correction in $M_h$ to the
Shapiro time delay. Finally, the last term in this
equation describes the first order in $a$ effect from rotation;
this effect can be traced back to the dragging of inertial frames.

From equation [\ref{eq:1}], the relative time
delay of two signals, emitted at point $r_e$, traveling in opposite directions
around the black hole and arriving at point $r_o$ is given by
\begin{equation}
\Delta t_s \equiv [t^-(r_e,d)+t^-(r_o,d)]-[t^+(r_e,d)+t^+(r_o,d)]
     = \frac{16aM_h}{d} \, .
\label{eq:2}
\end{equation}
To obtain an order of magnitude estimate
of the relative time delay for photon orbits slightly off the equatorial plane,
we approximate the impact parameter by $d \approx a_p (1-e) \cos{i}$,
with $a_p$, $e$ and $i$ the semi-major axis, eccentricity and
inclination of the pulsar's orbit, respectively. Equation~[\ref{eq:2}] then
yields
\begin{equation}
\Delta t_s \approx
4\mu\hbox{s}\left(\frac{a}{M_h}\right)\left(\frac{M_h}{20M_\odot}\right)^2
\left(\frac{5R_\odot}{a_p}\right)\left(\frac{0.5}{1-e}\right)
\left(\frac{\cos{89.5^o}}{\cos{i}}\right) \, .
\label{eq:3}
\end{equation}
This order of magnitude estimate indicates that
only orbital inclinations $i\ge 89^o$ and companion masses
$M_h > 10 M\odot$ would yield a potentially detectable effect
($\Delta t_s \sim \hbox{few}\, \mu\hbox{s}$).
Black hole masses of $\sim 10 M_\odot$ are believed to be typical for NS--BH
binaries
(\cite{phinney91,narayan91}).

Another possibility to increase the detectability of this effect is
to decrease the characteristic orbital separation $a_p$. However, this decreases
the lifetime for the binary to spiral in and merge due to the emission
of gravitational radiation.  The lifetime to merging, $\tau_m$, 
is approximately given by
\begin{equation}
\tau_m \equiv \frac{P_b}{\dot P_b}
 \approx 4\times 10^8 \hbox{yr} \left(\frac{a_p}{R_\odot}\right)^4
\left(\frac{M_\odot^3}{M_p M_h M_t}\right)F(e)^{-1} \, ,
\label{eq:4}
\end{equation}
with
$$
F(e) = \left(1+\frac{73}{24}e^2+\frac{37}{96}e^4\right)
\left(1-e^2\right)^{-7/2} \, , \nonumber
$$
$M_p$ the pulsar's mass, and $M_t = M_p+M_h$.
Thus, because of the $\tau_m \propto a_p^4$ dependence, any decrease of $a_p$
or increase of $e$, to
bring up $\Delta t_s$ in [\ref{eq:3}] above the detectability level,
will make the binary unlikely to be observable at that separation.

Our study was carried out under the following assumptions which favor
detectability:
{\it (i)} the line of sight is perpendicular to the rotational axis of
the companion, so the gravitational time delay is maximized, {\it (ii)}
the longitude of periastron is $\omega = 180^o$ (superior conjunction) to
yield a complete anti-symmetric (delay-acceleration) effect, and {\it (iii)}
orbital parameters are selected such that the lifetime to merging never
becomes shorter than $10^7$ yrs.
Time delay differences, $\Delta t_s$, are then computed by means of the
ray-tracing technique.  A bundle of photons
is integrated back from the point of observation, $r_o$, towards
the binary orbital plane using a fourth-order Runge-Kutta integrator
with variable time-step.
Photons passing within a distance $h \le 10^{-4} a_p$ of the
pulsar's orbit are tagged, and their time delay is computed by
comparing their propagation time with that from a flat space-time trajectory.
Approximately, $20,000$ photons are required to obtain a sufficiently complete
sample of time delays throughout the orbital phase.
In order to guarantee that photons in the bundle passing within
a distance $h$ are recorded, the integration step in the vicinity of the
orbital plane is chosen to be $< h$.
Once the time delays are recorded, the data are
interpolated to a uniform orbital phase grid.
Finally, difference in time delays are computed by subtracting
the time delays of pair of points along the pulsar's orbit
that otherwise would have identical travel time in flat space.
That is, the data are folded about the orbital phase of closest approach
and then subtracted.
For orbits in the equatorial plane, the above procedure is entirely equivalent
to measuring the last term in [\ref{eq:1}].
Furthermore, since this is a differential measurement, we have verified that
the value measured of $\Delta t_s$ does not depend on the distance to the
observer $r_o$ provided $r_o \gg a_p$ (see [\ref{eq:1}]).

\begin{figure}
\plotone{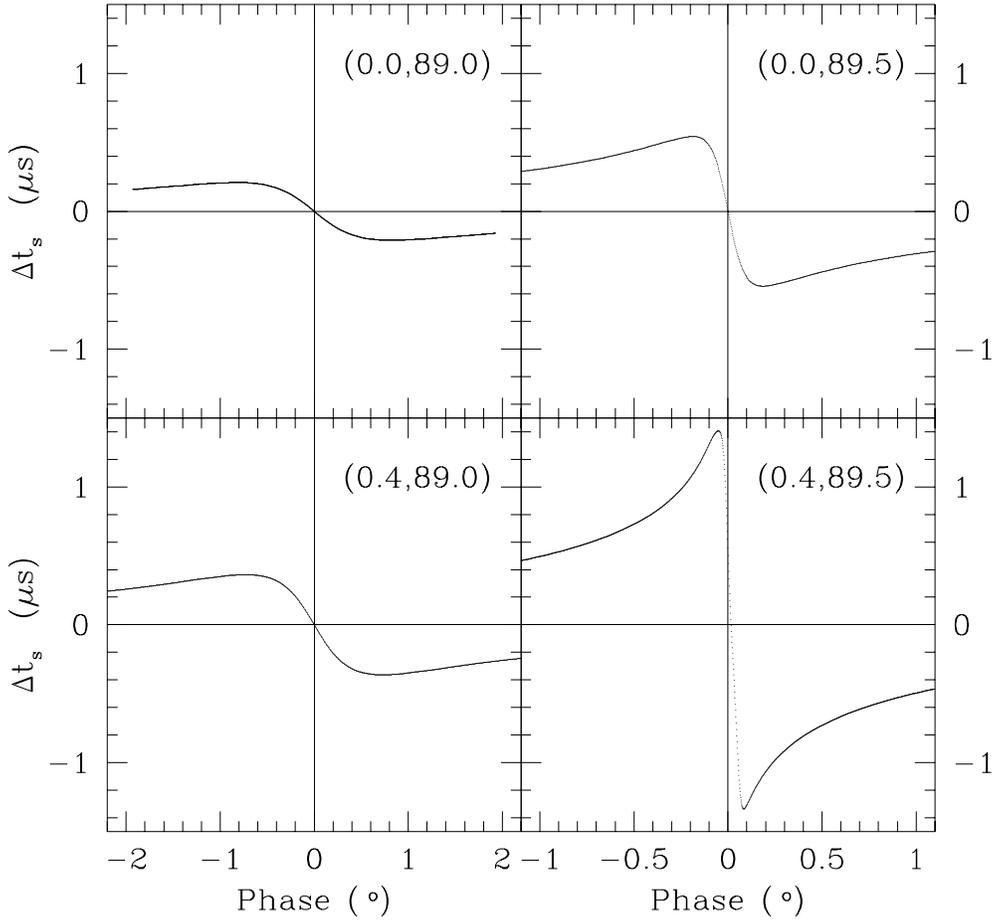}
\caption{Differences in arrival times, $\Delta t_s\, (\mu$s),
for PSR--BH systems with $M_h = 20M_\odot$, $M_p = 1.4M_\odot$,
$a = M_h$ and $a_p = 5R_\odot$.
Values in parentheses denote the respective
eccentricities and orbital inclinations
corresponding to cases 1, 2, 7 and 8 in table 1.
\label{fig1}}
\end{figure}

To verify the validity of our approach,
we conducted, in addition to standard convergence tests, a null test
with a non-rotating black hole and a test with orbital inclination
$i=90^o$. The latter test produced results that differed
by $\le 2\%$ from those obtained with equation [\ref{eq:3}].
Table~\ref{tbl-1} summarizes the results for our best case scenarios,
with $M_p = 1.4M_\odot$, $M_h = 20M_\odot$ and $a = M_h$.
The last column reports the maximum time delay $\Delta t_s$ when
$a_p$, $e$ and $i$ are varied.
A natural consequence of considering high orbital inclinations is that
the scalings of $\Delta t_s$ with $a/M_h$ and $M_h$,
implied by equation [\ref{eq:3}], are to some extent preserved.
We have conducted a series of simulations varying
$M_h$ and $a$ which suggest that the scaling of equation [\ref{eq:3}]
with respect to those quantities is indeed present for sufficiently
large orbital inclinations. Regarding $a_p$, $e$ and $i$,
the situation is not as clear. These parameters determine the
photon distance ($d$) of closest approach to the black hole;
thus, differences between the results in Table~\ref{tbl-1}
and those obtained from [\ref{eq:3}] provide an indication
of the extent to which the ``off the equatorial plane" approximation
breaks down, namely $d \approx a_p (1-e) \cos{i}$.
In particular, the scaling is absent for the cases with very large
inclinations $> 89.5^o$ and/or small periastron separations $< R_\odot$, for
which higher order effects in $a$ become important.
Figure~\ref{fig1} depicts the typical outcomes of our simulations.
They correspond to cases 1, 2, 7 and 8 in Table~\ref{tbl-1}.
In Figure \ref{fig1}, positive and negative values imply delays and
accelerations, respectively.

Our calculations demonstrate that, for reasonably compact and eccentric
orbits, rotation of a sufficiently massive black hole
companion to the orbiting pulsar leads to $\sim\mu$s--order departures
from pulse arrival times predicted for a Shapiro delay in the case
of a non--rotating body. Clearly, it is not practical to consider
tighter orbits, because, even if they were plausible on the grounds of
stellar evolution, their lifetimes against the
gravitational radiation would be extremely short.

A detectability of $\sim\mu$s--level effects in pulse timing residuals
has been demonstrated for several millisecond pulsars (e.g.
\cite{kaspi94,camilo94}).
Unfortunately, these very rapidly rotating neutron stars are the products of
the kind of binary evolution that is 
expected to not yield black hole companions
(\cite{phinney94}). Recent calculations (\cite{lipunov94})
indicate that the most typical NS--BH binaries,
resulting from the evolution of massive stars, would contain normal
``slow'' pulsars, in which case the practical timing accuracy
is rarely better than 0.2--1.0 ms. In addition, such pulsars are
short--lived ($\sim 10^7$ yr) and are expected to move in relatively wide,
long--period orbits.

Less probable, but not implausible evolutionary scenarios which could
produce compact NS--BH binaries, possibly containing
rapidly spinning ``recycled'' pulsars, involve an accretion induced
collapse of one of the stellar binary companions to a black hole
(\cite{narayan91,lipunov94}). For such systems,
sufficiently extreme orbital parameters (Table 1), and a possible
microsecond--level precision of the 
pulse timing measurements could create more favorable
conditions for a detection of the effects discussed in this paper. For example,
PSR B1534+12, one of the NS--NS binaries mentioned
above, which has been timed for more than four years, allows a detection
of orbital phase--dependent effects in timing residuals with a $\sim 1\mu$s
accuracy (\cite{arzou95}). For NS--BH binaries timable with such precision,
our study implies that, among other applications, a measurement of
the rotational contribution to the Shapiro delay could
be used to determine the angular momentum of the black hole companion.

This research was supported in part through NSF grants PHY-9601413 and
PHY-9357219 (NYI) to PL and AST-9619552 to AW.

\clearpage

\begin{deluxetable}{crcccccc}
\footnotesize
\tablecaption{Arrival time delay differences of pulses from binary
pulsars with rotating black hole companions for the
pulsar mass $M_p = 1.4M_\odot$, the black hole mass $M_h = 20M_\odot$,
and the black hole angular momentum per unit mass $a = M_h$.
\label{tbl-1}}
\tablewidth{0pt}
\tablehead{
\colhead{Case} &
\colhead{$i\, (\,^o)$} &
\colhead{$a_p\, (R_\odot)$} &
\colhead{$P_b$ (days)} &
\colhead{$e$} &
\colhead{$\tau_m\, (10^8$ years)} &
\colhead{$\Delta t_s\, (\mu$s)} &
}
\startdata
 1 &  89.0 &  5 &   0.2803   & 0.0  &   4.19   &   0.203  \nl
 2 &  89.0 &  5 &   0.2803   & 0.4  &   1.52   &   0.358  \nl
 3 &  89.0 &  5 &   0.2803   & 0.8  &   0.04   &   1.556  \nl
 4 &  89.0 & 10 &   0.7928   & 0.0  &   67.1   &   0.099  \nl
 5 &  89.0 & 10 &   0.7928   & 0.4  &   24.4   &   0.168  \nl
 6 &  89.0 & 10 &   0.7928   & 0.8  &   0.61   &   0.564  \nl
 7 &  89.5 &  5 &   0.2803   & 0.0  &   4.19   &   0.535  \nl
 8 &  89.5 &  5 &   0.2803   & 0.4  &   1.52   &   1.396  \nl
 9 &  89.5 &  5 &   0.2803   & 0.8  &   0.04   &   2.511\tablenotemark{a}  \nl
10 &  89.5 & 10 &   0.7928   & 0.0  &   67.1   &   0.219  \nl
11 &  89.5 & 10 &   0.7928   & 0.4  &   24.4   &   0.407  \nl
12 &  89.5 & 10 &   0.7928   & 0.8  &   0.61   &   2.091\tablenotemark{a}  \nl
\enddata
\tablenotetext{a}{Values taken at $0.1^o$ from superior conjunction}
\end{deluxetable}

\clearpage

\end{document}